# Nanoscale Control over Magnetic Light-Matter Interactions

*Benoît Reynier,[1] Eric Charron,[1] Obren Markovic,[1] Bruno Gallas,[1] Alban Ferrier,[2,3] Sébastien Bidault[4] and Mathieu Mivelle[1*]*

[1] Sorbonne Université, Centre National de la Recherche Scientifique, Institut des NanoSciences de Paris, 75005 Paris, France

[2] Chimie ParisTech, Paris Sciences & Lettres University, Centre National de la Recherche Scientifique, Institut de Recherche de Chimie Paris, 75005 Paris, France

[3] Faculté des Sciences et Ingénierie, Sorbonne Université, UFR 933, Paris 75005, France

[4] Institut Langevin, ESPCI Paris, Université PSL, CNRS, 75005 Paris, France

[*]mathieu.mivelle@sorbonne-universite.fr

Light-matter interactions are frequently perceived as predominantly influenced by the electric optical field, with the magnetic component of light often overlooked. Nonetheless, the magnetic aspect plays a pivotal role in various optical processes, including chiral light-matter interactions, photon-avalanching, and forbidden photochemistry, underscoring the significance of manipulating magnetic processes in optical phenomena. Here, we explore the ability to control the magnetic light and matter interactions at the nanoscale. In particular, we demonstrate experimentally, using a plasmonic nanostructure, the transfer of energy from the optical magnetic field to a nanoparticle, thanks to the deep subwavelength magnetic confinement allowed by our nano-antenna. This control is made possible by the particular design of our plasmonic nanostructure, which has been optimized to spatially separate the electric and magnetic fields of the localized plasmon.

Furthermore, by studying the spontaneous emission from the Lanthanide-ions doped nanoparticle, we observe that the optical field distributions are not spatially correlated with the electric and magnetic near-field quantum environments of this antenna, which seemingly contradicts the reciprocity theorem. We demonstrate that this counter-intuitive observation is in fact, the result of the different optical paths followed by the excitation and emission of the ions, which forbids a direct application of that theorem.

# 1. Introduction

Controlling light-matter interactions at the nanoscale has brought about transformative advancements across various scientific domains. Applications span from high sensitivity in diagnostic platforms for biochemistry[1] to precise nanoparticle-mediated medical therapies,[2] increased catalytic efficiency in chemistry,[3] and the exploration of exotic light-matter interaction processes in optical physics.[4] Despite the substantial progress achieved this far, the focus has been set on manipulating the electric light, with the magnetic component often being neglected. Indeed, traditionally, light-matter interactions are thought to be primarily influenced by the electric optical field, disregarding the significance of the magnetic counterpart. Nevertheless, the magnetic component assumes a critical role in numerous optical processes, including chiral light-matter interactions,[5] ultrasensitive detection,[6] enhancement of Raman optical activity,[7] photon-avalanching,[8] or forbidden photochemistry.[9] Hence, the manipulation of magnetic processes becomes crucial. Recent investigations have successfully demonstrated the control over specific interactions involving magnetic light and matter, in particular spontaneous emission[10, 11, 12, 13, 14-16] and stimulated excitation[16, 17] mediated by magnetic transition dipoles in Lanthanide ions. While the control and enhancement of magnetic luminescence was successfully investigated at scales both larger (through the use of metallic layers as mirrors)[10, 16] and smaller than the wavelength of light (thanks to dielectric[11, 15] and plasmonic nanostructures[13, 14]) by locally tuning the magnetic local density of states, the manipulation of magnetic stimulated excitation was limited to diffraction-limited dimensions, using either a focused azimuthally polarized laser beam[17] or stationary waves.[16] In this study, we demonstrate the nanoscale control over both stimulated excitation and spontaneous emission in $Eu^{3+}$ ions thanks to a plasmonic nano-antenna.

Here, we have designed a plasmonic nano-antenna with the aim of confining and enhancing the optical magnetic field at subwavelength scales. Thanks to the properties of light in the near-field, this magnetic hotspot is spatially isolated from its electric counterpart, providing a purely magnetic nanosource of light. This plasmonic nanostructure is placed at the apex of a near-field scanning optical microscope (NSOM), enabling the magnetic hot spot to be deterministically positioned close to a Lanthanide ion-doped nanoparticle. Through this deterministic coupling between the nano-antenna carrying the nanosource of magnetic light and the nanoparticle, we demonstrate the optical excitation of the latter at subwavelength scales by transferring the energy from the optical magnetic field to the nanomaterial under consideration. This interaction also enables us to map the nanoscale distribution of electric and magnetic fields of the localized plasmon generated by the plasmonic nanostructure, establishing the strongly subwavelength nature of the magnetic confinement. Also, by studying the spontaneous emission from the doped nanoparticle, we observe that the optical field distributions are not spatially correlated with the electric and magnetic near-field quantum environments of this antenna, which seemingly contradicts the reciprocity theorem.[18] We demonstrate that this counter-intuitive observation is in fact the result of the different optical paths followed by the excitation and emission of the nanoparticle, which forbids a direct application of that theorem in our experimental configuration.

## 2. Results

The optical nano-antenna used in this study comprises an aluminum nanodisk measuring 50 nm in thickness and 550 nm in diameter, as depicted in **Figure 1**. Fabricated at the end of a pulled optical fiber tip (see Supporting Information), the nano-antenna serves as a local probe for an NSOM when affixed to a tuning fork. This integration offers two crucial advantages. Firstly, it facilitates direct excitation of the nano-antenna through the optical fiber (Figure 1) by injecting the laser beam directly into the fiber core, allowing its propagation to the tip and nanodisk. Moreover, the NSOM's three-dimensional nanometric manipulation of the tip enables precise control of the antenna's position relative to the sample of interest. In our case, the sample consists of yttrium oxide ($Y_2O_3$) nanoparticles with an approximate diameter of 150 nm, doped with trivalent europium ($Eu^{3+}$) ions (Figure 1 and Supporting Information). These ions, of particular interest for this investigation, exhibit purely electric (ED) or magnetic (MD) dipolar transitions both at excitation and emission, as illustrated in the partial band diagram in Figure 1.[19] The study leverages these specific europium properties to explore the coupling between magnetic or electric optical fields and matter at the nanoscale.

The dimensions of the nanostructure are meticulously chosen to ensure that under excitation at wavelengths $\lambda_{exc}^{MD}$=527.5 nm (MD: $^7F_0$→$^5D_1$) and $\lambda_{exc}^{ED}$=532 nm (ED: $^7F_1$→$^5D_1$), the magnetic and electric fields do not spatially overlap in the localized plasmon of the antenna. Additionally, the plasmonic nanostructure is designed to confine the magnetic field in its core, as illustrated in **Figures 2**a-c. Experimentally, the excitation wavelength selection targeting either the MD or ED transition of $Eu^{3+}$ ions is achieved by finely filtering a supercontinuum laser source. The NSOM's nanopositioning and feedback capabilities enable the approach and scanning of the plasmonic nanostructure within a few nanometers and in the plane of the doped nanoparticles, facilitating their near-field excitation by localized plasmon for all antenna-particle positions. Furthermore, by collecting the luminescence signal emanating from the MD ($^5D_0 \rightarrow {}^7F_1$, $\lambda_{em}^{MD}$ = 593 nm) and ED ($^5D_0 \rightarrow {}^7F_2$, $\lambda_{em}^{ED}$ = 611 nm) transitions of $Eu^{3+}$ in the C2 site (spectrum and band diagram in Figure 1), at each nanoparticle position, the distribution of the exciting fields, as well as the local density of magnetic (MLDOS) and electric (ELDOS) optical states, can be imaged.

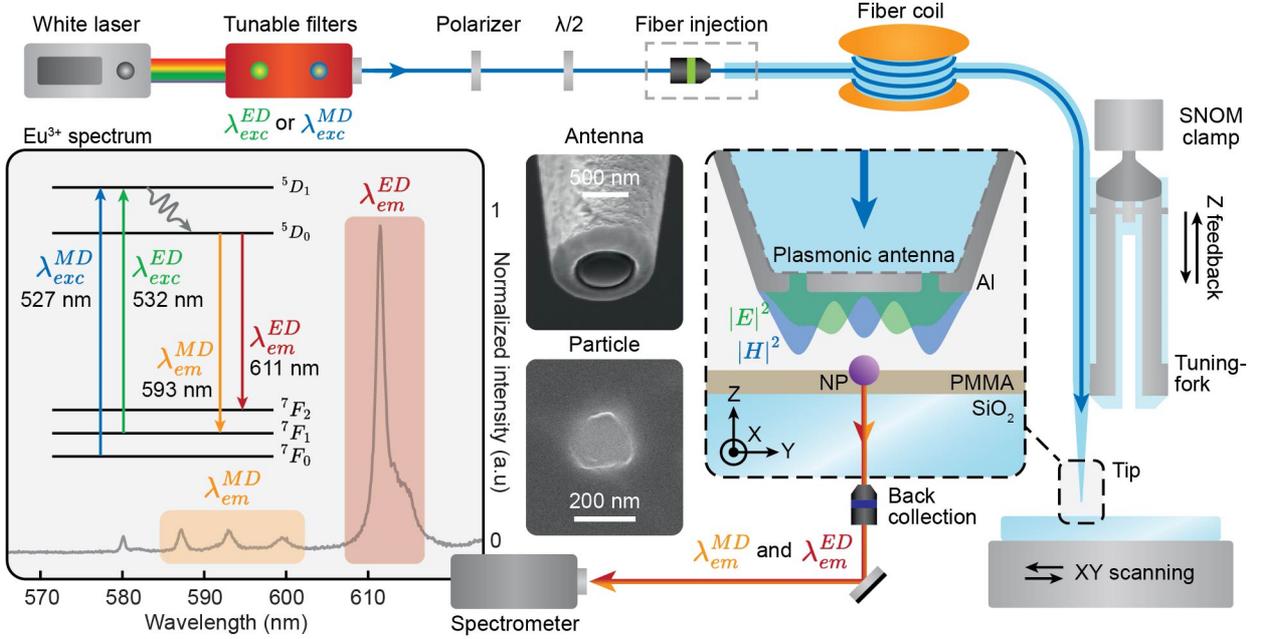

**Figure 1**. Illustration of the experimental configuration. An aluminum nanodisk serves as a plasmonic nano-antenna and is precisely fabricated at the end of an NSOM fibered tip. This tip is glued to a tuning fork, and via a feedback loop mechanism, the antenna's position can be deterministically controlled within a few nanometers from the sample, ensuring precise manipulation in all three spatial dimensions. The plasmonic nanostructure is excited by a supercontinuum laser, filtered using a series of interference filters to isolate a specific wavelength range with a 2 nm bandwidth. The laser beam, rendered linearly polarized, is injected into the optical fiber supporting the tip and antenna, resulting in the optical excitation of the latter. The localized plasmon generated by the antenna is used to excite $Eu^{3+}$:$Y_2O_3$ nanoparticles, which are deposited on a glass substrate. SEM images of an antenna and a nanoparticle are shown. Luminescence emitted by the nanoparticles is collected using an immersion objective (×100, NA=1.3) from the substrate side and measured with a spectrometer. The inset provides the emission spectrum of europium ions in the $Y_2O_3$ matrix, along with the partial band diagram of these emitters.

Figures 2a and b depict the theoretical distributions (see Supporting Information for simulation details) of electric (at $\lambda_{exc}^{ED}$ = 532 nm) and magnetic (at $\lambda_{exc}^{MD}$ = 527.5 nm) field intensities within a plane situated inside the nanoparticle beneath the plasmonic antenna excited by linear polarization (inset in Figure 2b). A noticeable distinction is observed in the spatial profiles of these fields, characteristic of a cavity mode. Specifically, the electric field manifests a two-lobes pattern on the outer regions of the plasmonic nanodisk, whereas the magnetic field displays a three-lobes motif, with one centrally positioned within the disk and two outer lobes in the groove separating the disk from the remainder of the fibered tip. In Figures 2c and d, the normalized experimental luminescence signal is presented when scanning a $Eu^{3+}$ ion-doped nanoparticle in the vicinity of the plasmonic nano-antenna, as outlined in Figure 1, when exciting the ED and MD transitions at wavelengths $\lambda_{exc}^{ED}$ = 532 nm and $\lambda_{exc}^{MD}$ = 527.5 nm, respectively.

The comparison between numerical simulations and experimental outcomes reveals a remarkable agreement. Employing the laser source at the wavelength corresponding to the MD transition effectively leads to the excitation of europium ions through the magnetic field of the localized plasmon. Conversely, employing the wavelength associated with the ED transition results in ion excitation via the electric field. This not only validates the capability to selectively excite matter through the electric or magnetic field of a localized plasmon but also underscores the potential for imaging the full distribution of the field components of the light at deep subwavelength scales in the proximity of a plasmonic antenna. This is achieved through the scanning capability of the NSOM and the proportionality of the luminescence signal to the field intensities.

Furthermore, Figures 2e and f present line cuts of the theoretical and experimental distributions, respectively, as depicted in Figures 2a-d, with green or blue lines corresponding to the electric and magnetic field distributions. The spatial decoupling of the optical fields is evident, and again a very good agreement is observed between the simulated field distributions and the spatially dependent luminescence signals. Notably, the magnetic field is prominently localized at the center of the antenna within a subwavelength area. Specifically, the experimental curves in Figure 2f illustrate that the optical magnetic field is confined to a region of only 130 nm. These experiments unequivocally demonstrate the targeted coupling of the magnetic field of a localized plasmon to a nanoparticle at deep subwavelength scales. Moreover, the NSOM capability not only enables selective excitation of matter by the magnetic field but also facilitates the full-scale imaging of light at these spatial dimensions.

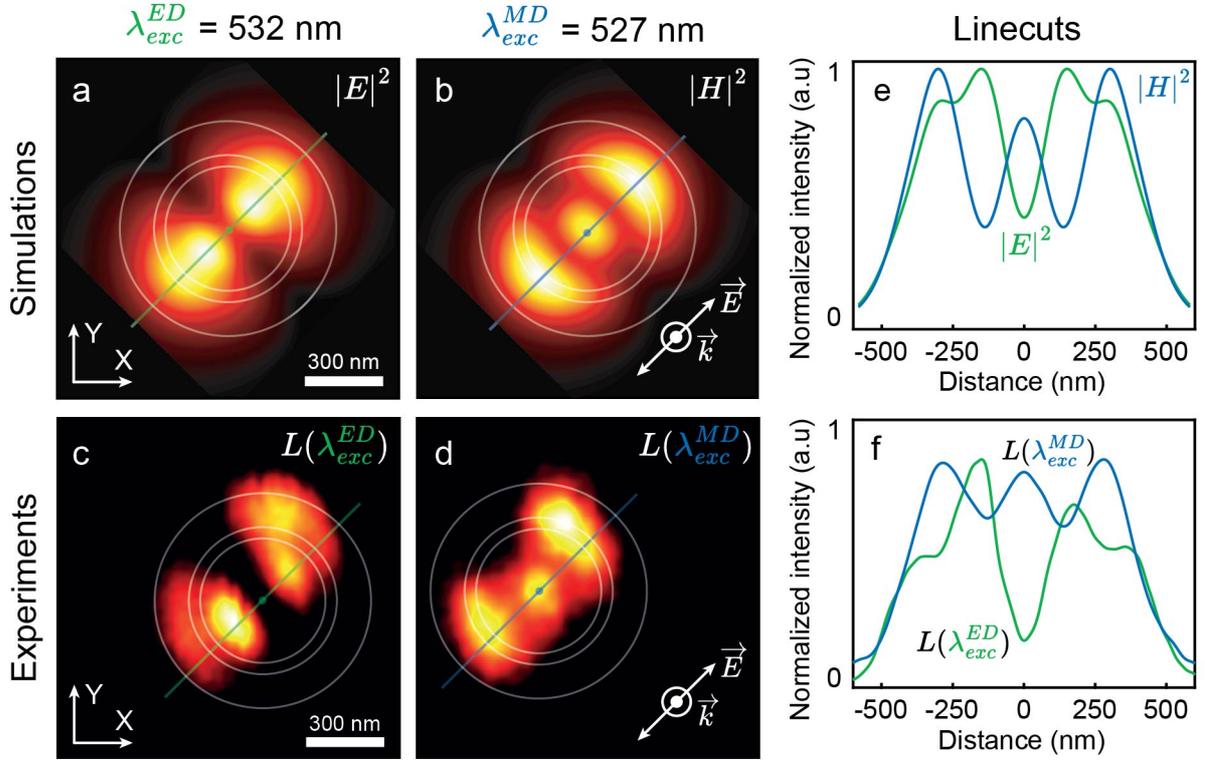

**Figure 2**. Near-field excitation of the doped nanoparticle by the localized plasmon. Theoretical representations of a) the integrated electric field intensity distribution (at $\lambda_{exc}^{ED}$ = 532 nm) and b) the integrated magnetic field intensity distribution (at $\lambda_{exc}^{MD}$ = 527.5 nm) in an XY plane beneath the aluminum nanodisk at a distance Z of 100 nm. The linear polarization of the excitation is as indicated in inset b). Normalized luminescence distributions collected during the scan of the plasmonic nanostructure excited at c) the wavelength of the ED transition ($\lambda_{exc}^{ED}$ = 532 nm) and d) the MD transition ($\lambda_{exc}^{MD}$ = 527.5 nm). The linear polarization used is shown in inset d). Line cuts are presented for e) theoretical field intensities and f) experimental luminescence signals obtained from the distributions in (a, b) and (c, d), respectively. Green lines correspond to line cuts of electric signals, while blue lines correspond to the magnetic counterpart.

The partial band diagram presented in Figure 1 illustrates that $Eu^{3+}$ ions exhibit ED and MD transitions during both excitation and emission processes.[16] Notably, since emission transitions originate from the same energy level, they can serve as a means to characterize the relative MLDOS and ELDOS within a photonic environment.[14, 15] Consequently, by examining the luminescence emission, particularly the ratios between the emitted photons in each transition for every position of the plasmonic nano-antenna, the distribution of electric and magnetic LDOS under this structure can be traced with nanoscale precision (see Supporting Information for theoretical and experimental description of LDOS calculations).

**Figure 3** illustrates the ELDOS and MLDOS distributions beneath the antenna. Given our experimental approach's precise control over all nanoscale light-matter interactions—both electric and magnetic—during excitation and emission, the LDOS can be computed in various manners. Indeed, as the LDOS represents the quantum environment of a photonic structure and is associated with the spontaneous emission of the quantum emitters, it is independent of the optical excitation. Therefore, the LDOS can be calculated for both electric or magnetic excitations. With this in mind, for each excitation, luminescence emission is distinguished between the contributions of ED ($^5D_0 \rightarrow {}^7F_2$, $\lambda_{em}^{ED}$ = 611 nm) and MD ($^5D_0 \rightarrow {}^7F_1$, $\lambda_{em}^{MD}$ = 593 nm) transitions.

In Figures 3a and b, the luminescence distribution of these two transitions is shown for excitation via the magnetic transition ($^7F_0 \rightarrow {}^5D_1$) at $\lambda_{exc}^{MD}$ = 527.5 nm. Similarly, Figures 3c and d display luminescence emission at wavelengths of 611 nm and 593 nm, respectively, for particle excitation via the electric transition ($^7F_1 \rightarrow {}^5D_1$) at $\lambda_{exc}^{ED}$ = 532 nm. Subsequently, based on these luminescence mappings, it becomes feasible to compute the electric and magnetic LDOS for each of the electric and magnetic excitations, as depicted in Figures 3e-h. Figures 3e and g illustrate the ELDOS for magnetic and electric excitation, respectively, while Figures 3f and h depict the MLDOS for these same magnetic and electric excitations.

Here, several observations can be made. Firstly, as anticipated, the LDOS exhibit similar behavior irrespective of the field component used to excite the doped nanoparticle. However, it is surprisingly observed that the ELDOS increases at the center of the antenna, concomitant with a decrease in the MLDOS. This observation is intriguing, given that upon antenna excitation, it is the optical magnetic field that experiences an increase at the center of the antenna, not the electric field. According to the reciprocity theorem, these two processes should be symmetrical.[18] The expectation is that an increase in an optical field should result in a corresponding increase in its LDOS, but this appears not to be the case in this particular scenario. It is crucial to note here that it is the ability to map the distribution of fields and LDOS at the nanoscale that enables the observation of this apparent inconsistency.

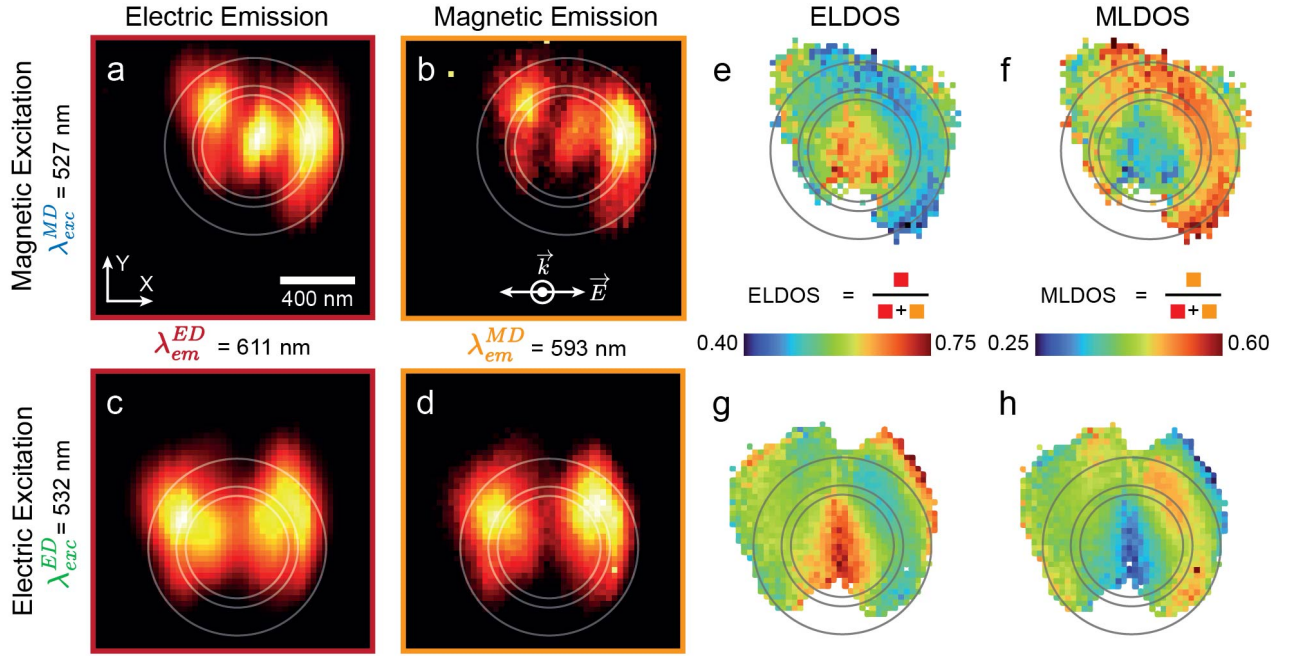

**Figure 3**. Spatial distributions of electric and magnetic LDOS. Luminescence distributions from a $Eu^{3+}$ ion-doped nanoparticle induced by localized plasmon excitation during nanodisk scanning are presented for excitation via a,b) the MD transition ($^7F_0 \rightarrow {}^5D_1$) at $\lambda_{exc}^{MD}$ = 527.5 nm and c,d) the ED transition ($^7F_1 \rightarrow {}^5D_1$) at $\lambda_{exc}^{ED}$ = 532 nm. These luminescence distributions are further segregated into emission contributions via a,c) ED transition ($^5D_0 \rightarrow {}^7F_2$, $\lambda_{em}^{ED}$ = 611 nm) and b,d) MD transition ($^5D_0 \rightarrow {}^7F_1$, $\lambda_{em}^{MD}$ = 593 nm). Subsequently, the e,g) ELDOS and f,h) MLDOS are plotted for e,f) magnetic and g,h) electric excitations, respectively.

The pertinent question here is why this discrepancy exists. The answer lies in the distinct optical paths followed during laser excitation and signal collection. The reciprocity theorem, which asserts equality when emitter and detector positions are swapped, can be extended to the radiative decay rate of a quantum source and the associated exciting optical field if the source emission and the excitation wave follow the same optical path. In other words, if the emission and excitation share the same wavevectors with opposite signs.[18] In our experimental setup, excitation is performed through the optical fiber, while the luminescence is collected by the immersion objective. Consequently, the optical paths are diametrically opposite, thereby explaining the absence of reciprocity in this context.

Numerical simulations confirm this hypothesis. In **Figure 4**, a comparison is presented between the theoretical electric and magnetic LDOS beneath the plasmonic antenna, and the experimental data. Figures 4a-d depict the distributions of electric (Figures 4a,c) and magnetic (Figures 4b,d) radiative LDOS in two different directions—along the tip direction (positive Zs, Figures 4a,b) and towards the substrate (negative Zs, Figures 4c,d).

To simplify the problem, the antenna is modeled as an infinite layer of aluminum, disregarding the tip. However, for a better analogy with the experimental setup, the collection of the luminescence signal in the substrate takes into account the numerical aperture of the microscope objective (Figure 1). Moreover, the signal collected in the upper part accounts only for the energy radiated directly above the nanodisk (see Supporting Information for simulation details). As one can see, the results are striking: the spatial distribution of the magnetic and electric LDOS are completely reversed. While the ELDOS is maximum in the center of the antenna for radiation towards the substrate, it is minimal for radiation towards the upper part, and vice versa for the MLDOS. A comparison with the experimental results shown in Figures 3e-h and 4e,f indicates that the experimental distributions align well with theoretical LDOS radiated towards the substrate. These results shed new light on the limits of the reciprocity theorem, especially when considering the magnetic component of light, which, to our knowledge, has never been discussed in the literature.

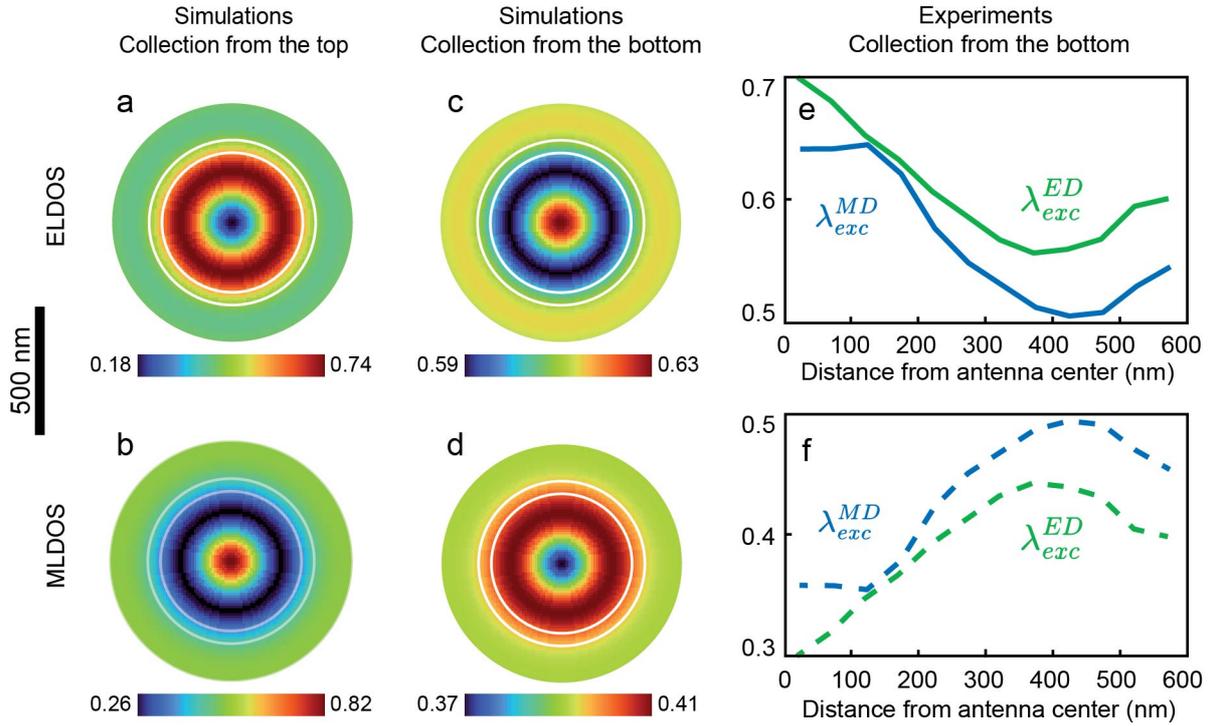

**Figure 4**. Comparative analysis of the radiative magnetic and electric LDOS for spontaneous emission towards the substrate or in the opposite direction. The distributions of a,c) electric and b,d) magnetic radiative LDOS are presented for radiation in the direction of a,b) the NSOM tip and c,d) the substrate. Experimental e) ELDOS and f) MLDOS for electric (green lines) and magnetic (blue lines) excitation, starting from the center of the nanodisk to its edge. Experimental curves are obtained by averaging the LDOS provided in figures 3e-h in a circular pattern from the center of the nanodisk to increase the signal-to-noise ratio.

## 3. Conclusion

In summary, our study leveraged a photonic nanostructure to demonstrate the selective excitation of a solid-state nanoparticle by the optical magnetic field of a localized plasmon. The nanoscale confinement of the magnetic field by the nano-antenna revealed that this transfer of energy occurs at strongly subwavelength scales. Through precise targeting of excited optical transitions, we achieved nanoscale imaging of all electric and magnetic components of the localized plasmon within the nanostructure. The versatility in selecting the exciting optical field, coupled with the ability to choose luminescence emission via electric and magnetic transitions of the doped particle, facilitated the imaging of the spatial distribution of electric and magnetic LDOS that are manipulated by the antenna at a subwavelength scale as well.

Furthermore, based on these LDOS distributions, our analysis demonstrated that the reciprocity theorem, applied to the magnetic field of light, could not be applied here due to the different optical paths taken by the optical excitation and the collected luminescence emission. Notably, this study of the reciprocity theorem applied to the magnetic field represents, to our knowledge, the first experimental report on this subject.

The manipulation of the coupling between magnetic light and matter at the nanoscale, particularly through plasmonic nanostructures, unveils promising prospects across various research domains, such as chiral light-matter interactions,[5] photochemistry,[9] manipulation of magnetic processes,[20] and new schemes in quantum computing[21] or nonlinear processes,[8] among others.

**Supporting Information**

Supporting Information is available from the Wiley Online Library or from the author.

**Acknowledgements**

This work is supported by the Agence Nationale de la Recherche (ANR-20-CE09-0031-01, ANR-22-CE09-0027-04 and ANR-23-ERCC-0005), the Institut de Physique du CNRS (Tremplin@INP 2020).

The table of contents illustrates a localized plasmon wave, where the bright regions depict the optical electric field, and the dark areas represent the magnetic field of light. The deterministic excitation of a nanoparticle by the optical magnetic field is demonstrated, highlighting the coupling between the dark side of light and matter.

B.R. Author1, E.C. Author 2, O.M. Author 3, B.G. Author 4, A.F. Author 5, S.B. Author 6, M.M. Corresponding Author*

**Nanoscale Control over Magnetic Light-Matter Interactions**

ToC figure

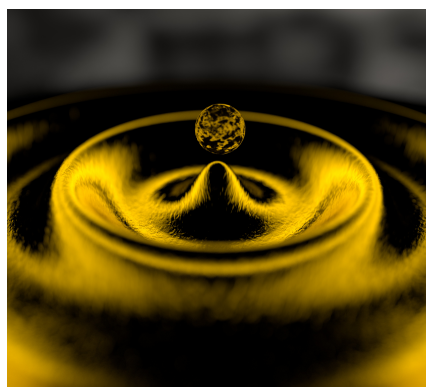

# Bibliography

[1] a) D. Punj, M. Mivelle, S. B. Moparthi, T. S. van Zanten, H. Rigneault, N. F. van Hulst, M. F. García-Parajó, J. Wenger, *Nat. Nanotechnol.* **2013**; b) P. M. Winkler, R. Regmi, V. Flauraud, Vol. 9, 2018.

[2] a) D. P. O'Neal, L. R. Hirsch, N. J. Halas, J. D. Payne, J. L. West, *Cancer letters* **2004**, 209, 171; b) P. Fortina, L. J. Kricka, D. J. Graves, J. Park, T. Hyslop, F. Tam, N. Halas, S. Surrey, S. A. Waldman, *Trends Biotechnol.* **2007**, 25, 145.

[3] a) R. Grisel, K.-J. Weststrate, A. Gluhoi, B. E. Nieuwenhuys, *Gold Bulletin* **2002**, 35, 39; b) R. Sardar, A. M. Funston, P. Mulvaney, R. W. Murray, *Langmuir* **2009**, 25, 13840.

[4] a) T. Taminiau, F. Stefani, F. Segerink, N. Van Hulst, *Nat. Photonics* **2008**, 2, 234; b) M. L. Juan, M. Righini, R. Quidant, *Nat. Photonics* **2011**, 5, 349; c) D. Akinwande, C. Huyghebaert, C.-H. Wang, M. I. Serna, S. Goossens, L.-J. Li, H.-S. P. Wong, F. H. Koppens, *Nature* **2019**, 573, 507; d) J. J. Baumberg, J. Aizpurua, M. H. Mikkelsen, D. R. Smith, *Nat. Mater.* **2019**, 18, 668; e) B. Yang, G. Chen, A. Ghafoor, Y. Zhang, Y. Zhang, Y. Zhang, Y. Luo, J. Yang, V. Sandoghdar, J. Aizpurua, *Nat. Photonics* **2020**, 14, 693.

[5] Y. Tang, A. E. Cohen, *Phys. Rev. Lett.* **2010**, 104, 163901.

[6] Z. Xi, H. Urbach, *Phys. Rev. Lett.* **2017**, 119, 053902.

[7] T. Wu, X. Zhang, R. Wang, X. Zhang, *J. Phys. Chem. C* **2016**, 120, 14795.

[8] C. Lee, E. Z. Xu, Y. Liu, A. Teitelboim, K. Yao, A. Fernandez-Bravo, A. M. Kotulska, S. H. Nam, Y. D. Suh, A. Bednarkiewicz, *Nature* **2021**, 589, 230.

[9] A. Manjavacas, R. Fenollosa, I. Rodriguez, M. C. Jiménez, M. A. Miranda, F. Meseguer, *Journal of Materials Chemistry C* **2017**, 5, 11824.

[10] a) N. Noginova, Y. Barnakov, H. Li, M. Noginov, *Opt. Express* **2009**, 17, 10767; b) S. Karaveli, R. Zia, *Physical review letters* **2011**, 106, 193004. c) T. H. Taminiau, S. Karaveli, N. F. van Hulst, R. Zia, *Nat. Commun.* **2012**, 3, 979; d) S. M. Hein, H. Giessen, *Phys. Rev. Lett.* **2013**, 111, 026803; e) L. Aigouy, A. Cazé, P. Gredin, M. Mortier, R. Carminati, *Phys. Rev. Lett.* **2014**, 113, 076101; f) F. T. Rabouw, P. T. Prins, D. J. Norris, *Nano. Lett.* **2016**, 16, 7254.

[11] a) B. Rolly, B. Bebey, S. Bidault, B. Stout, N. Bonod, *Phys. Rev. B* **2012**, 85, 245432; b) T. Feng, W. Zhang, Z. Liang, Y. Xu, A. E. Miroshnichenko, *ACS Photonics* **2017**, 5, 678; c) T. Feng, Y. Xu, Z. Liang, W. Zhang, *Opt. Lett.* **2016**, 41, 5011; d) A. Vaskin, S. Mashhadi, M. Steinert, K. E. Chong, D. Keene, S. Nanz, A. Abass, E. Rusak, D.-Y. Choi, I. Fernandez-Corbaton, *Nano. Lett.* **2019**; e) P. R. Wiecha, C. Majorel, C. Girard, A. Arbouet, B. Masenelli, O. Boisron, A. Lecestre, G. Larrieu, V. Paillard, A. Cuche, *Appl. Opt.* **2019**, 58, 1682; f) X. Cheng, X. Zhuo, R. Jiang, Z. G. Wang, J. Wang, H. Q. Lin, *Advanced Optical Materials* **2021**, 9, 2002212; g) H. Sugimoto, M. Fujii, *ACS Photonics* **2021**, 8, 1794; h) Y. Brûlé, P. Wiecha, A. Cuche, V. Paillard, G. C. des Francs, *Opt. Express* **2022**, 30, 20360; i) A. Bashiri, A. Vaskin, K. Tanaka, M. Steinert, T. Pertsch, I. Staude, *ACS Nano* **2023**.

[12] D. G. Baranov, R. S. Savelev, S. V. Li, A. E. Krasnok, A. Alù, *Laser & Photonics Reviews* **2017**, 11, 1600268.

[13] a) R. Hussain, S. S. Kruk, C. E. Bonner, M. A. Noginov, I. Staude, Y. S. Kivshar, N. Noginova, D. N. Neshev, *Opt. Lett.* **2015**, 40, 1659; b) M. Mivelle, T. Grosjean, G. W. Burr, U. C. Fischer, M. F. Garcia-Parajo, *ACS Photonics* **2015**, 2, 1071; c) B. Choi, M. Iwanaga, Y. Sugimoto, K. Sakoda, H. T. Miyazaki, *Nano. Lett.* **2016**, 16, 5191;

[14] C. Ernandes, H.-J. Lin, M. Mortier, P. Gredin, M. Mivelle, L. Aigouy, *Nano. Lett.* **2018**, 18, 5098.

[15] M. Sanz-Paz, C. Ernandes, J. U. Esparza, G. W. Burr, N. F. van Hulst, A. s. Maitre, L. Aigouy, T. Gacoin, N. Bonod, M. F. Garcia-Parajo, S. Bidault, M. Mivelle, *Nano. Lett.* **2018**, 18, 3481.

[16]     B. Reynier, E. Charron, O. Markovic, X. Yang, B. Gallas, A. Ferrier, S. Bidault, M. Mivelle, *arXiv preprint arXiv:2301.06345* **2023**.

[17]     M. Kasperczyk, S. Person, D. Ananias, L. D. Carlos, L. Novotny, *Phys. Rev. Lett.* **2015**, 114, 163903.

[18]     a) R. Carminati, M. Nieto-Vesperinas, J.-J. Greffet, *JOSA A* **1998**, 15, 706; b) P. Bharadwaj, B. Deutsch, L. Novotny, *Advances in Optics and Photonics* **2009**, 1, 438; c) D. Cao, A. Cazé, M. Calabrese, R. Pierrat, N. Bardou, S. Collin, R. Carminati, V. Krachmalnicoff, Y. De Wilde, *Acs Photonics* **2015**, 2, 189; d) H. Benisty, J.-J. Greffet, P. Lalanne, *Introduction to nanophotonics*, Oxford university press, **2022**.

[19]     N. Chang, J. Gruber, *The Journal of Chemical Physics* **1964**, 41, 3227.

[20]     D. Bossini, V. I. Belotelov, A. K. Zvezdin, A. N. Kalish, A. V. Kimel, *Acs Photonics* **2016**, 3, 1385.

[21]     D. Serrano, S. K. Kuppusamy, B. Heinrich, O. Fuhr, D. Hunger, M. Ruben, P. Goldner, *Nature* **2022**, 603, 241.

**Supporting information**

**Nanoscale Control over Magnetic Light-Matter Interactions**

*Benoît Reynier,[1] Eric Charron,[1] Obren Markovic,[1] Bruno Gallas,[1] Alban Ferrier,[2,3] Sébastien Bidault[4] and Mathieu Mivelle[1]\**

[1] Sorbonne Université, Centre National de la Recherche Scientifique, Institut des NanoSciences de Paris, 75005 Paris, France
[2] Chimie ParisTech, Paris Sciences & Lettres University, Centre National de la Recherche Scientifique, Institut de Recherche de Chimie Paris, 75005 Paris, France
[3] Faculté des Sciences et Ingénierie, Sorbonne Université, UFR 933, Paris 75005, France
[4] Institut Langevin, ESPCI Paris, Université PSL, CNRS, 75005 Paris, France
\*mathieu.mivelle@sorbonne-universite.fr

## 1. $Eu^{3+}$ ion-doped $Y_2O_3$ nanoparticles

The europium ions $Eu^{3+}$ offer the potential for excitation through various transitions. In particular, the transition at $\lambda_{exc}^{MD} = 527,5$ nm and the transition at $\lambda_{exc}^{ED} = 532$ nm have been shown to be mediated by magnetic and electric transition dipoles, respectively. Figures S1a and b display the excitation spectra achieved by scanning the excitation wavelength in increments of 1nm and by collecting the electric and magnetic dipole emission peaks at $\lambda_{em}^{ED} = 611$ nm and $\lambda_{em}^{MD} = 593$ nm.

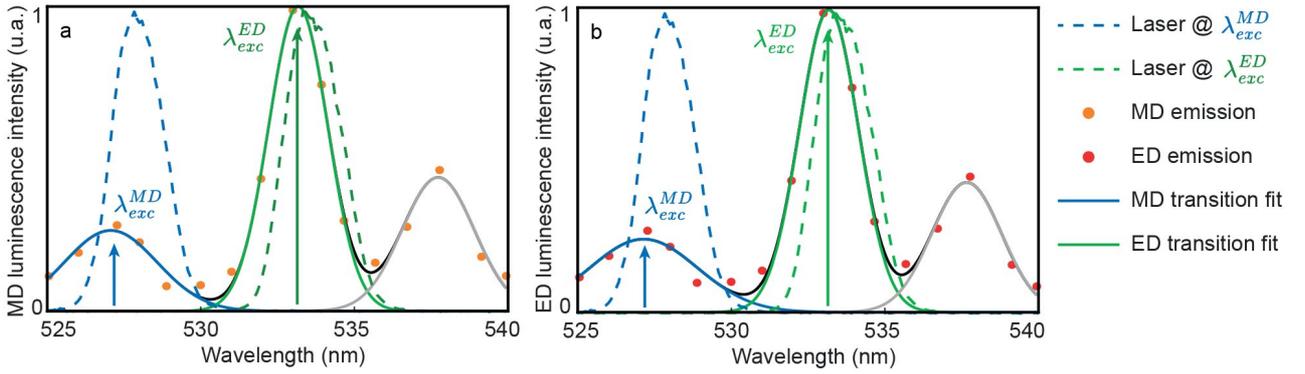

**Figure S1.** a) Excitation spectrum of $Eu^{3+}$-doped nanoparticles for a collection at $\lambda_{em}^{MD} = 593$ nm ($^5D_0 \rightarrow {}^7F_1$). b) Excitation spectrum of $Eu^{3+}$-doped nanoparticles for a collection at $\lambda_{em}^{ED} = 611$ nm ($^5D_0 \rightarrow {}^7F_2$). The laser lines used in the study to excite the electric and magnetic transitions are shown as green and blue dashed lines, respectively. The corresponding electric ($^7F_1 \rightarrow {}^5D_1$) and magnetic ($^7F_0 \rightarrow {}^5D_1$) transitions are shown as green and blue solid line fits. As we can see, the excitation peaks are independent of the emission channel.

## 2. Theoretical background: excitation study

The luminescence L at the emission wavelength $\lambda_i$ for the i transition (i = ED or MD), excited by the field A (with A the electric E or magnetic H optical field) can be defined as:

$$L(A, \lambda_i) = \sigma(A) \times |A|^2 \times \eta(\lambda_i) \times Q(\lambda_i), \quad (S1)$$

where, $\sigma(A)$ is the absorption cross-section, $|A|^2$ is the electric or the magnetic field intensity, $\eta(\lambda_i)$ and $Q(\lambda_i)$ are the collection efficiency and the quantum yield of the transition, respectively.

## 3. Theoretical background: calculation of the LDOS

Exploiting the electric ($^5D_0 \rightarrow {}^7F_2$) and magnetic ($^5D_0 \rightarrow {}^7F_1$) dipolar transitions emanating from the same excited level, the $Eu^{3+}$ ions serve as ideal candidates for probing the local quantum environment. Consequently, one can compute for each spectrum the corresponding electric $\beta^{ED}$ and magnetic $\beta^{MD}$ branching ratios using the following expressions:

$$\beta^{ED} = \frac{L^{ED}}{L^{ED}+L^{MD}} = 1 - \beta^{MD},$$

where $L^{MD}$ and $L^{ED}$ are the collected luminescence at $\lambda_{em}^{MD} = 593$ nm and $\lambda_{em}^{ED} = 611$ nm, respectively. Next, by comparing the calculated branching ratios in the presence of the plasmonic antenna ($\beta_{PA}$) to a reference situation without the antenna and the nanoparticle excited from the farfield ($\beta_0$), one can directly compute the relative variation of the Electric (or Magnetic) Local Density Of States, denoted as ELDOS (or MLDOS) using the formula:

$$\tilde{\rho}^{ED} = \frac{\rho_{PA}^{ED}/\rho_0^{ED}}{\rho_{PA}^{ED}/\rho_0^{ED} + \rho_{PA}^{MD}/\rho_0^{MD}} = \frac{\beta_{PA}^{ED}/\beta_0^{ED}}{\beta_{PA}^{ED}/\beta_0^{ED} + \beta_{PA}^{MD}/\beta_0^{MD}} = 1 - \tilde{\rho}^{MD}$$

ELDOS and MLDOS can also be calculated numerically, by knowing that:

$$\frac{\rho_{PA}}{\rho_0} = \frac{\Gamma_{PA}}{\Gamma_0} = \frac{P_{PA}}{P_0},$$

where $\Gamma_{PA}$ and $\Gamma_0$ are the radiate rates experienced for a dipole with the plasmonic antenna and without, respectively. $P_{PA}$ and $P_0$ are the corresponding numerically calculated radiated power (see section Methods for more details).

## 4. Additional luminescence images:

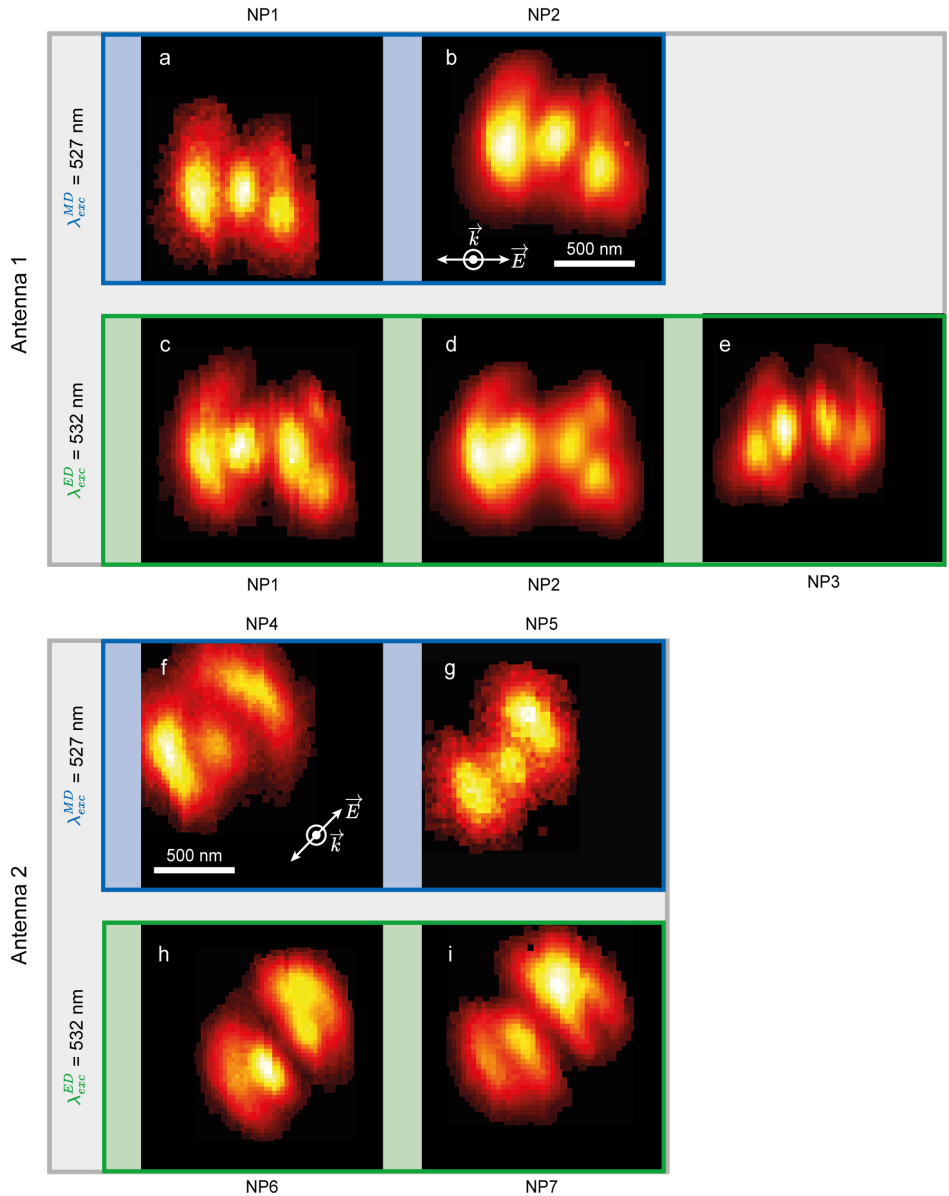

**Figure S2.** Additional Luminescence Images. Two different scanning probe tips featuring plasmonic antennas with the same dimensions, denoted as Antenna 1 (a-e) and Antenna 2 (f-i), were employed to excite different europium-doped nanoparticles both magnetic (a,b and f,g) and electric (c,d,e and h,i) wavelengths. All luminescence images are normalized and share the same size scale. a) and c) luminescence images have been done with the same nanoparticle (NP1), as well as b) and d) (NP2). Note that g) and h) provide the dataset showed in Figure 1. All measurements provide similar results that are in good agreement with the electric and magnetic fields simulated in the near-field of the antenna. Small discrepancies probably arise from the size and shape of the nanoparticle and from a potential tilt of the fiber impinging on the sample.

## 5. Methods

Plasmonic nanodisk fabrication:

The fabrication process for the nanostructured tips involves multiple sequential steps. Initially, an optical fiber is pulled using a P-2000 puller from Sutter to create the fibered tip. Subsequently, a layer of 120 nm aluminum is deposited around the perimeter of the fibered tip, primarily for focused ion beam (FIB) purposes. The tip is then precision-cut by a focused ion beam to achieve a core diameter of approximately 800 nm. Following this, a second layer of aluminum, 50 nm in thickness, is thermally evaporated onto the processed end section of the tip. Lastly, a circular groove is created using FIB to form a nanomirror with a diameter of 550 nm.

Nanoparticles synthesis:

2% Eu:$Y_2O_3$ nanoparticle of 150 ± 50 nm average diameter were prepared by homogeneous precipitation.[38] In this method, an aqueous nitrate solution of $Y(NO_3)_3 \cdot 6H_2O$ (99.9% pure, Alfa Aesar), $Eu(NO_3)_3 \cdot 6H_2O$ (99.99% pure, Reacton) was mixed with an aqueous urea solution ($CO(NH_2)_2$, > 99% pure, Sigma) in a Teflon reactor. The pH inside the reactor was then slowly increased during a 24 h thermal treatment at 85°C by the urea decomposition. The metal and urea concentrations were 7.5 mmol $L^{-1}$ and 3 mol $L^{-1}$ respectively. After cooling, a white precipitate of amorphous yttrium hydroxycarbonate ($Eu^{3+}$: $Y(OH)CO_3 \cdot n\, H_2O$) was collected by centrifugation and washed at least 3 times with water and absolute ethanol. That amorphous powder was then converted to highly crystalline Eu:$Y_2O_3$ nanoparticles by 24 h calcination treatment at 1000°C (rate of 3 °C $min^{-1}$). The body-centred cubic $Y_2O_3$ structure (Ia-3 space group) of the particles was confirmed by X-ray diffraction with no evidence for other parasitic phases.

Sample preparation:

$Eu^{3+}$ doped nanoparticle samples. After cleaning by sonic bath and plasma cleaner, a 110 nm layer of PMMA is deposited by spin-coating (3% weight - 4000 rpm) on glass coverslips then annealed at 180° for one minute to evaporate the excess solvent and homogenize the layer. To make the layer hydrophilic, the sample is once again treated with plasma cleaner, reducing the PMMA layer to 80 nm thickness. The doped nanoparticles are then deposited by spin-coating on the sample, and new annealing is performed at 180° for one minute to allow the nanoparticles to embed in the polymer.

Simulations:

The simulations were conducted using the finite difference time domain (FDTD) software Lumerical. The aluminum nano-antenna, positioned at the end of a fibered tip, features a diameter of 550 nm with a metal thickness of 50 nm. This metal structure is separated from the remainder of the tip by a 50 nm gap. The optical index of the glass fiber tip is 1.46, and the permittivity of aluminum employed in the simulation was determined experimentally using ellipsometry. Beneath the tip, a spherical $Y_2O_3$ nanoparticle with a diameter of 150 nm and an optical index of 1.94 is introduced. This nanoparticle is positioned on a glass substrate (index of 1.5) and is partially embedded in a 100 nm layer of PMMA (index of 1.46). The overall size of the simulation window is approximately 2x2x1.5 μm³, and the finest mesh, defining the most detailed parts, has a resolution of 5 nm.

For investigating electric and magnetic field distributions, a Gaussian beam was introduced into the fiber at 800 nm, directed towards the plasmonic antenna. Additionally, to consider the impact of nanoparticle size, an integration of electric and magnetic fields within the nanoparticle was performed for each position beneath the nano-antenna. This approach yielded a qualitatively simulated luminescence distribution. The electric and magnetic fields were calculated at $\lambda_{exc}^{MD}$ and $\lambda_{exc}^{ED}$, respectively. The results were normalized by the intensities of electric and magnetic fields of the incoming plane wave.

A simplified model was employed to analyze the ELDOS and MLDOS, as well as to examine the reciprocity theorem. Electric and magnetic dipoles were positioned beneath a 50 nm thick 2D infinite aluminum layer featuring a 550 nm diameter disk with a 50 nm gap to model the plasmonic antenna. The considered electric and magnetic dipoles were emitting at $\lambda_{em}^{ED}$ and $\lambda_{em}^{MD}$, respectively. For each dipole orientation, the emitted power was collected from below, considering the experimental Numerical Aperture (NA), and from above, specifically just on top of the antenna center, at the origin position. To mimic the isotropic orientation of the $Eu^{3+}$ ions, the results for each dipole oriented along X, Y, and Z were averaged. Similar simulations were conducted without the aluminum layer for the radiated power references (labelled by 0 0 subscript). The position of the dipoles is then scanned below the antenna to provide the maps of the ELDOS and MLDOS. Finally, a convolution by the nanoparticle size is performed and provides the ELDOS and MLDOS experienced aby the emitters s featured in figure 4 of the main text.

Setup:

Excitation of $Eu^{3+}$-doped nanoparticles is performed by a supercontinuum laser (NKT Photonics K90-110-10), filtered by a combination of interference filters (Semrock BrightLine FF01-532/18-25 and Spectrolight FWS-B-F06), in order to reduce the spectral bandwidth to 2 nm while maintaining high laser power. First, the excitation light is finely polarized and injected into the optical fiber. Then, the end of the fiber coil is welded to the optical fiber supporting the nano-antenna. The optical near-field microscope (NT-MDT-Integra) is placed on an inverted microscope (Olympus IX73), and the tip supporting the nano-antenna is glued on a tuning fork vibrating at a frequency of 32kHz. The approach and the feedback loop of the tip in the near field are performed by monitoring the phase of the oscillation of the tuning fork (oscillation below 1 nm[1, 2]). Next, the tip is aligned on the center of an oil immersion objective (Olympus PLN 100x Oil Immersion, NA 1.30), and the particle is scanned under it thanks to a piezoelectric stage (Piezoconcept), allowing a nanometric displacement. Then, the luminescence is collected from below and sent to a spectrometer (Sol Instruments MS5204i) after high-pass filter (Semrock BrightLine FF552-Di01-25x36). The luminescence spectra are then measured with a CCD camera (Andor iDus 401 CCD) for each particle-antenna position, leading to hyperspectral images.

**Bibliography**

(1) van Hulst, N. F.; Veerman, J.-A.; Garcıa-Parajó, M. F.; Kuipers, L. Analysis of individual (macro) molecules and proteins using near-field optics. The Journal of Chemical Physics **2000**, 112 (18), 7799-7810.

(2) Koopman, M.; De Bakker, B.; Garcia-Parajo, M.; Van Hulst, N. Shear force imaging of soft samples in liquid using a diving bell concept. Appl. Phys. Lett. **2003**, 83 (24), 5083-5085.